\documentclass[twocolumn,superscriptaddress,aps,pre,amssymb,amsfonts,amsmath]{revtex4-1}
\usepackage{bm}
\usepackage{hyperref}
\usepackage{graphicx}
\usepackage{xcolor}
\newcommand{\rev}[1]{\textcolor{black}{{#1}}}

\begin{document}

\title{Random-field Ising model criticality in a glass-forming liquid}%

\author{Benjamin Guiselin}%

\email{benjamin.guiselin@umontpellier.fr}

\affiliation{Laboratoire Charles Coulomb (L2C), Universit\'e de Montpellier, CNRS, 34095 Montpellier, France}

\author{Ludovic Berthier}%

\affiliation{Laboratoire Charles Coulomb (L2C), Universit\'e de Montpellier, CNRS, 34095 Montpellier, France}

\affiliation{Department of Chemistry, University of Cambridge, Lensfield Road, Cambridge CB2 1EW, United Kingdom}

\author{Gilles Tarjus}%
\affiliation{LPTMC, CNRS-UMR 7600, Sorbonne Universit\'e, 4 Pl. Jussieu, F-75005 Paris, France}

\date{\today}%

\begin{abstract}
We use computer simulations to investigate the extended phase diagram of a supercooled liquid linearly coupled to a quenched reference configuration.
An extensive finite-size scaling analysis demonstrates the existence of a random-field Ising model (RFIM) critical point and of a first-order transition line, \rev{in agreement with recent field-theoretical approaches.} The dynamics in the vicinity of this critical point resembles the peculiar activated scaling of RFIM-like systems, and the overlap autocorrelation displays a logarithmic stretching. \rev{Our study demonstrates RFIM criticality in the thermodynamic limit for a three-dimensional supercooled liquid at equilibrium.}
\end{abstract}
\maketitle

\section{Introduction}
What is the best starting point for a proper theoretical description of glass formation in supercooled liquids? The fact that this question remains hotly debated reflects the difficulty to provide a definite, and then widely accepted, resolution of the problem~\cite{berthier2011theoretical,tarjus2011overview,chandler2010dynamics}. One strong candidate ascribes the slowing down of relaxation to properties of the free energy landscape and to the presence of an underlying thermodynamic transition to an ideal glass phase~\cite{lubchenko2007theory}. This transition is unreachable, as it lies below the experimental glass transition $T_g$, but is nonetheless supposed to control glass formation in real materials. This theoretical approach, the random first-order transition (RFOT) theory~\cite{kirkpatrick1989scaling}, takes its strength from the exact analytical solution of glass-forming liquids in the limit of infinite dimensions of space, which realizes exactly the predicted scenario at a mean-field level~\cite{kurchan2012exact,parisi2020theory}. Going from infinite to three dimensions is however a nontrivial qualitative leap, because spatial fluctuations are expected to play a key role and the very concepts of metastable states and free energy landscape become ill-defined. 

What remains of the mean-field scenario in three dimensions? The dynamical (mode-coupling-like~\cite{gotze2008complex}) transition found at the mean-field level can at best survive as a crossover in finite dimensions~\cite{kirkpatrick1989scaling,lubchenko2003barrier,rizzo2014long}, due to thermally activated processes, and its detection is always subject to interpretations. As for the putative RFOT at $T_K<T_g$, it is not directly testable, even with efficient swap Monte Carlo algorithms~\cite{ninarello2017models}. The mean-field/RFOT description puts the focus on an order parameter, the similarity or overlap between liquid configurations, and on its statistics. Following the well-established statistical mechanical formalism for phase transitions, one is then led to consider the role of specific boundary conditions and associated length scales~\cite{bouchaud2004adam} or, alternatively, of pinning fields and applied sources~\cite{monasson1995structural,franz1997phase,cammarota2012ideal}. In this context, it is found that, at least at the mean-field level, applying a nonzero source $\epsilon$ linearly coupled to the overlap order parameter generates a line of first-order transition emanating from the RFOT at $(T_K,~\epsilon=0)$ and terminating in a critical point at a higher temperature $(T_c>T_K,~\epsilon_c)$~\cite{franz1997phase,franz1998effective}. \rev{Recent field-theoretical arguments beyond mean-field~\cite{franz2013universality,biroli2014random} predict} that this critical point should be in the universality class of the random-field Ising model (RFIM). The goal of the present work is to test whether this prediction is realized in a realistic three-dimensional glass-forming liquid~\footnote{Numerical and analytical arguments have previously been given for simple lattice (plaquette) glass models~\cite{jack2016phase,biroli2016role}.}. \rev{Several previous attempts} exist~\cite{franz1998effective,cardenas1998glass,cardenas1999constrained,cammarota2010phase,berthier2013overlap,berthier2015evidence}, but \rev{their conclusions have been mostly qualitative because of the impossibility to work at a low enough temperature or because of much too small system sizes. We make here a qualitative, decisive step forward by being able to study the proper range of temperatures as well as large system sizes (an order of magnitude larger than previous numerical investigations) allowing for an extensive finite-size study of the transition, which is the standard (but highly demanding) tool to analyze phase transitions.} 
Furthermore, we characterize the nature of the \rev{slowing down of relaxation around the disordered critical point, which has never been done before.} This allows us to establish, as well as possible using atomistic simulations, that the terminal critical point is in the universality class of the RFIM. \rev{Our work demonstrates that a nontrivial piece} of the mean-field scenario is present in the phase diagram of finite-dimensional glass-forming liquids. \rev{This represents an additional physical application of the RFIM universality class, indeed an important topic for statistical mechanics studies of disordered systems.}

\section{Methods}
We consider a three-dimensional atomistic model glass-former that we study through state-of-the-art simulation techniques, including the recently developed swap algorithm~\cite{ninarello2017models,berthier2019efficient} that allows us to equilibrate liquid configurations down to the conventional glass transition temperature $T_g$, umbrella sampling~\cite{torrie1977monte,frenkel2001understanding} and reweighting techniques~\cite{challa1988gaussian} to properly sample rare configurations, and isoconfigurational ensemble to obtain a better statistics for the dynamics~\cite{widmer2004reproducible}. We focus on the overlap between a configuration ${\mathbf r}^N = \{r^{(i)}, i=1,\dots, N \}$ of $N$ atoms in equilibrium at temperature $T$ and a quenched reference configuration ${\mathbf r}_0^N$ equilibrated at a temperature $T_0$: $\hat{Q}[\mathbf r^N ;\mathbf r_0^N] = N^{-1} \sum_{i,j}w(|r^{(i)} - r_0^{(j)}|/a)$, where $w(x)$ is a strictly positive window function of width unity such that $w(0)=1$, $w(+\infty)=0$, and $a$ is a small length accounting for thermal vibrations around the reference configuration.

Thermodynamic fluctuations are characterized by the free energy cost to maintain the overlap $\hat{Q}$ at a given value $Q$, 
\begin{equation}
V(Q\vert T; \mathbf r_0^N, T_0)=-\frac {T}{N} \ln\int \mathrm{d}\mathbf r^N \frac{e^{-\beta \mathcal H[\mathbf r^N]}}{\mathcal Z(T)}\delta(\hat{Q}[\mathbf r^N; \mathbf r_0^N]-Q)
\label{eqn:V_FP}
\end{equation}
where $\beta=(k_BT)^{-1}$ (the Boltzmann constant is set to unity), $\mathcal H$ the liquid hamiltonian and $\mathcal{Z}(T)$ the partition function.
This free energy is obtained from the probability distribution of the overlap $\mathcal P(Q\vert T; \mathbf r_0^N, T_0)$, which is the argument of the logarithm in Eq.~(\ref{eqn:V_FP}). This is a random variable as it depends on the reference configuration $\mathbf r_0^N$, which is a source of quenched disorder. The average over $\mathbf r_0^N$ (taken with a Boltzmann distribution at temperature $T_0$) yields $V(Q\vert T;T_0)=~\overline{V(Q\vert T; \mathbf r_0^N, T_0)}$, which is called the Franz-Parisi potential~\cite{franz1995recipes,franz1997phase,franz1998effective}.

The dynamics near the critical point located at ($T_c$,~$\epsilon_c$) is investigated through the equilibrium overlap autocorrelation function 
\begin{equation}
C(t|\epsilon,T;\mathbf r^N_0,T_0)=\frac{\langle \delta \hat{Q}(t)\delta \hat{Q}(0) \rangle_\epsilon}{\langle \delta\hat{Q}(0)^2 \rangle_\epsilon},
\label{eqn:CQQ}
\end{equation}
where $\delta\hat{Q}=\hat{Q}- \langle \hat{Q} \rangle_\epsilon$ and $\langle \cdot \rangle_\epsilon$ denotes a thermal average at temperature $T$ in the presence of the applied source $\epsilon$, such that the liquid Hamiltonian is now $\mathcal H_\epsilon[\mathbf r^N; \mathbf r_0^N]=\mathcal H[\mathbf r^N]-N\epsilon \hat{Q}[\mathbf r^N; \mathbf r_0^N]$. The above correlation function is again a random function, through the dependence on the reference configuration. 

A severe obstacle that has hampered numerical studies of the putative critical point in the extended ($T$,~$\epsilon$) phase diagram is that when $T_0=T$ the critical point is expected at a temperature $T_c$ at which the relaxation time of the liquid is already so large that conventional simulation techniques are barely able to equilibrate the system at $\epsilon=0$. We have solved this problem by using the swap algorithm that allows an equilibration of the continuously polydisperse liquid mixture under consideration (see the Supplemental Information~\cite{SI}) much below what is attainable by standard methods~\cite{ninarello2017models,berthier2019efficient}. To give an idea, present-day molecular dynamics simulations equilibrate the model down to $T\approx 0.1$, which is near the mode-coupling crossover ($T_\mathrm{mct}=0.095$) whereas the swap algorithm allows equilibration down to $T\approx 0.055<T_g$. To characterize the critical point we have therefore chosen a low temperature $T_0=0.06 \lesssim T_g$ for sampling the equilibrium reference configurations, which has the prime merit of significantly increasing the critical temperature  $T_c(T_0)$ without 
altering its universality class~\cite{franz2013universality,biroli2014random,franz1997phase,franz1998effective}. \rev{We have also investigated whether the critical point persists when $T_0=T$, and we provide strong evidence that it does.}


We perform extensive computer simulations to study a wide range of system sizes, $N=300$, 600, 1200, 2400 at number density $\rho=1$. To perform the disorder average, we consider up to 28 different reference configurations. More details can be found in the SI~\cite{SI}. 

\begin{figure}
\includegraphics[width=8.5cm]{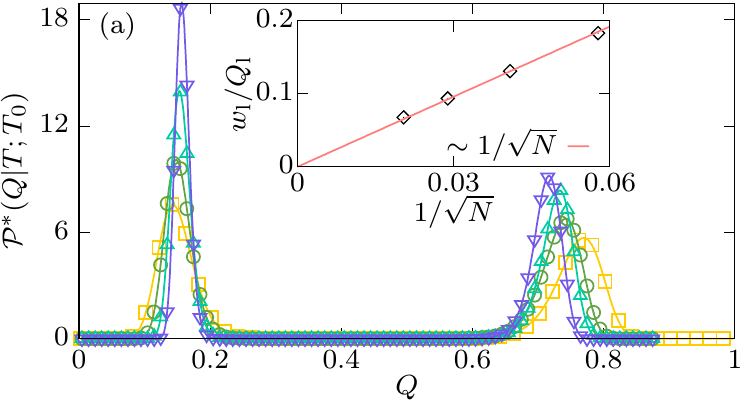}
\includegraphics[width=8.5cm]{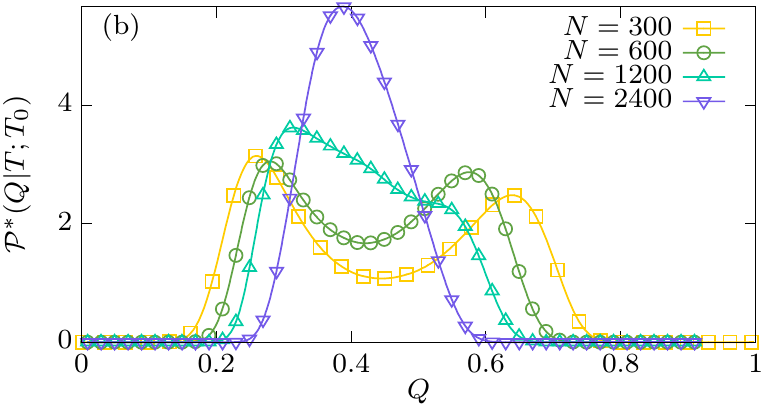}
\includegraphics[width=4.27cm]{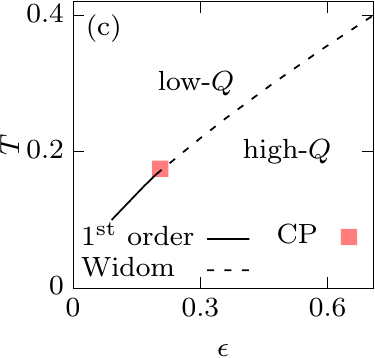}
\includegraphics[width=4.27cm]{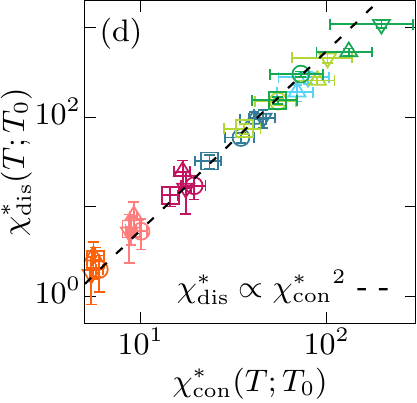}
\caption{(a,b) Overlap probability distribution  $\mathcal P^*(Q\vert T;T_0)$  as a function of system size $N$ below [$T=0.15$ in (a)] and above [$T=0.30$ in (b)] the critical point. The inset of (a) shows the half-width $w_\mathrm{l}$ of the low-overlap peak rescaled by the peak position $Q_\mathrm{l}$ as a function of $1/\sqrt{N}$. (c) ($T$,~$\epsilon$) phase diagram showing a first-order transition line at low temperature and a Widom line at high temperature, separated by a critical point (symbol). (d) Disconnected versus connected susceptibilities with a quadratic fit (dashed line); the symbols are as in panel (b), with the colors now denoting the different temperatures. Uncertainties are computed with the jacknife method~\cite{newman1999monte}.}
\label{fig:phase_diagram}
\end{figure}

\section{Finite-size scaling analysis}
We first present evidence for the presence of a transition line in the ($T$, $\epsilon$) diagram separating a low-overlap from a high-overlap phase. Operationally, we use a method developed to study systems in the presence of quenched disorder when, contrary to the standard RFIM, there is no $Z_2$ inversion symmetry~\cite{vink2008colloid,vink2010finite}. We compute the thermal susceptibility, $\chi_T(\epsilon,T;\mathbf r_0^N,T_0)=N\beta(\langle \hat Q^2 \rangle_\epsilon-\langle\hat Q\rangle_\epsilon^2)$, for each reference configuration $\mathbf r_0^N$ and temperature $T$, and we determine the location of its maximum, $\epsilon^*(T;\mathbf r_0^N,T_0)$. We next follow the evolution of the system along the disorder-averaged line $\epsilon^*(T;T_0)=\overline{\epsilon^*(T;\mathbf r_0^N,T_0)}$.
The behavior of the probability distribution of the overlap along this line, $\mathcal P^*(Q\vert T;T_0)=\overline{\mathcal P(Q\vert \epsilon^*(T;T_0),T;\mathbf r_0^N,T_0)}$, is illustrated in Figs.~\ref{fig:phase_diagram}~(a,b). For a low enough temperature [$T=0.15$ in Fig.~\ref{fig:phase_diagram}~(a)] the probability is clearly bimodal and the width of the two well-separated peaks shrinks as $N$ increases. The width of the low-overlap peak rescaled by the peak position follows the expected $N^{-1/2}$ behavior~\cite{binder1984finite}. This is strong evidence for the presence of a first-order transition at low temperature when $N\to \infty$. The finite-size scaling (FSS) of additional quantities is provided in the SI and supports as well the existence of a transition in the thermodynamic limit~\cite{SI}. For higher temperatures [$T=0.30$ in Fig.~\ref{fig:phase_diagram}~(b)], the probability distribution is bimodal for the smallest system sizes, but becomes single-peaked for the largest systems (hence the need to consider large system sizes and perform finite-size analysis to avoid considerably overestimating $T_c$). This region corresponds to a ``Widom line'' that is the locus of the (finite) maximum of the susceptibility. As one lowers the temperature along this line, one expects to cross a critical point at which the susceptibility diverges and below which a first-order transition is encountered, see Fig.~\ref{fig:phase_diagram}~(c). The overlap distributions at $T_c$ evolve very much as the low-temperature ones in Fig.~\ref{fig:phase_diagram}~(a).

\begin{figure}
\includegraphics[width=4.27cm]{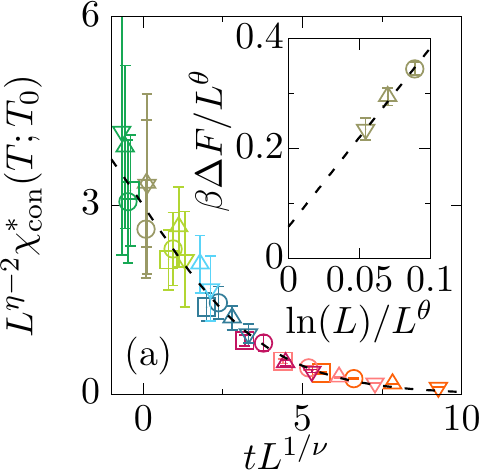}
\includegraphics[width=4.27cm]{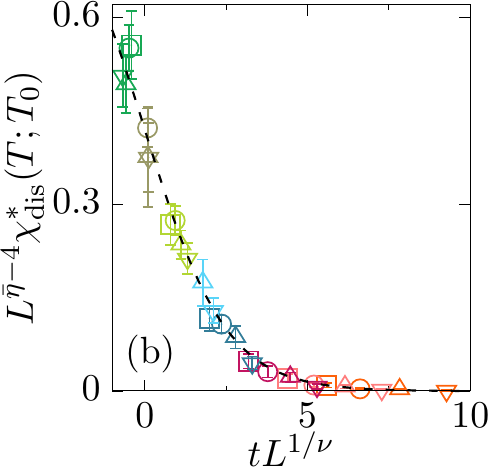}
\caption{(a) Connected susceptibility as a function of reduced temperature $t$ rescaled according to the FSS ansatz with $\eta=0.52$ and $\nu=1.37$. Data from all sizes collapse onto a master curve $\widetilde{\chi}_\mathrm{con}(x)$. The dashed line is a guide for the eye. (b) Equivalent plot for the disconnected susceptibility with $\bar \eta = 1.04$. In both panels, colors and symbols are as in Fig.~\ref{fig:phase_diagram}~(d). Data in grey for $t\approx 0$ are obtained via a temperature reweighting from data at $T=0.15$~\cite{SI}. The inset of panel (a) shows that the scaled free energy barrier at the critical point  $\beta\Delta F/ L^\theta$ approaches a positive nonzero value as $L\to\infty$ with a $\ln(L)/L^\theta$ behavior, where $\theta=1.49$.}
\label{fig:chi_FSS}
\end{figure}

For RFIM-like systems without inversion symmetry, ratios of cumulants of the order parameter are not a practical way to detect the critical point~\cite{vink2010finite}. Instead, to more precisely locate and characterize this critical point, we focus on the susceptibilities. Because of the quenched disorder associated with ${\bf r}_0^N$, and as in the case of the RFIM, one must consider two distinct susceptibilities, the connected one, $\chi_\mathrm{con}(\epsilon,T;T_0)= \overline{\chi_T(\epsilon,T;\mathbf r_0^N,T_0)}$, and the disconnected one, $\chi_\mathrm{dis}(\epsilon,T;T_0)= \beta N(\overline{\langle \hat{Q} \rangle_\epsilon^2} - \overline{\langle \hat{Q} \rangle_\epsilon}^2)$, which we evaluate at $\epsilon^*(T;T_0)$ for all temperatures and system sizes and then denote with a star. RFIM physics has a distinct signature in the behavior of these susceptibilities, because the disorder-induced fluctuations diverge much more strongly than thermal ones. As a result, for large but finite systems of linear size $L\propto N^{1/3}$ at the first-order transition and at the critical point~\footnote{Strictly speaking, the relation is not exact at the critical point. In a RFIM-like system of linear size $L$, $\chi_\mathrm{dis}\sim L^{4-\bar\eta}$ and $\chi_\mathrm{con}\sim L^{2-\eta}$, with $\bar\eta$ and $\eta$ the so-called anomalous dimensions. It turns out that for the $3$d RFIM, $\bar\eta \approx 2\eta$, so that indeed to a very good approximation $\chi_\mathrm{dis}\propto {\chi_\mathrm{con}}^2$~\cite{fytas2013universality,fytas2016efficient,tarjus2013critical}.}
\begin{equation} 
\chi_\mathrm{dis}^*(T;T_0)\propto \chi_\mathrm{con}^*(T;T_0)^2.
\label{eq:scaled-susceptibility}
\end{equation}
This relation is very well obeyed by our data, as shown in Fig.~\ref{fig:phase_diagram}~(d). When approaching the critical point from above along the Widom line, the susceptibilities should follow the FSS behavior, i.e., $\chi_\mathrm{con}^*(T;T_0)=L^{2-\eta} \widetilde \chi_\mathrm{con}(t L^{1/\nu})$ and $\chi_\mathrm{dis}^*(T;T_0)= L^{4-\bar\eta} \widetilde \chi_\mathrm{dis}(t L^{1/\nu})$, where $\eta$, $\bar\eta$ and $\nu$ are critical exponents, $t=(T/T_c-1)$ is the reduced temperature, and $\widetilde \chi_\mathrm{con}(x)$ and $\widetilde \chi_\mathrm{dis}(x)$ are scaling functions which are non-singular at $x=0$. In Fig.~\ref{fig:chi_FSS} we display the outcome of our FSS analysis where we have used the known values of the critical exponents for the $3d$ RFIM: $\eta\approx 0.52$, $\bar\eta\approx 1.04$ and $\nu\approx 1.37$~\cite{fytas2013universality}. The data collapse is excellent, with the critical point located at $T_c(T_0)\approx 0.167$ [which corresponds to $\epsilon_c(T_0)=\epsilon^*(T_c;T_0)\approx 0.20$]~\footnote{The best estimate of the critical temperature $T_c$ is found by minimizing the average quadratic difference between the rescaled data and the master curve~\cite{houdayer2004low,melchert2009autoscale}}. Hyperscaling violation also distinguishes the RFIM university class and implies that at the critical point the free energy barrier $\Delta F$ between the low-overlap and the high-overlap phases is not scale-invariant but instead grows as $\Delta F\sim \Upsilon L^\theta$, with $\theta\approx 1.49$ the temperature exponent~\cite{middleton2002three} and $\Upsilon$ finite and nonzero. 
To extract $\Delta F$, we measure the overlap distribution at $\epsilon^*(T;\mathbf r_0^N,T_0)$ for each individual sample. The inset of Fig.~\ref{fig:chi_FSS}~(a) shows that our data are compatible with a finite positive value of $\beta \Delta F/L^\theta\approx {0.08}$ in the thermodynamic limit~\cite{binder1982monte}. We provide additional FSS results in the SI~\cite{SI}.

\begin{figure}
\includegraphics[width=4.27cm]{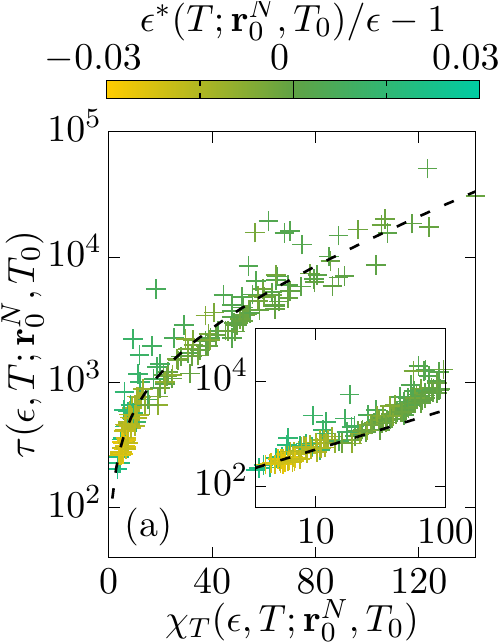}
\includegraphics[width=4.27cm]{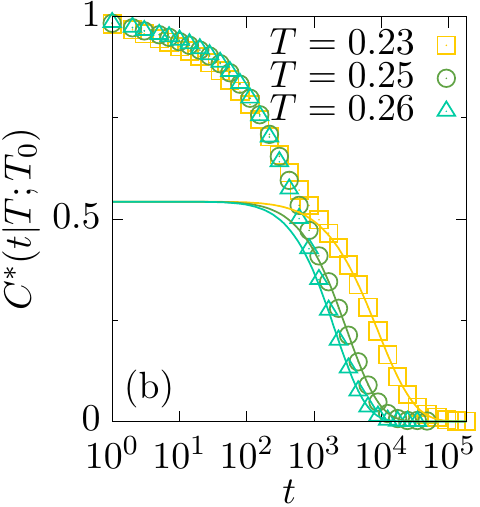}
\caption{(a) Relaxation time $\tau(\epsilon,T;\mathbf r_0^N,T_0)$ as a function of the thermal susceptibility $\chi_T(\epsilon,T;\mathbf r_0^N,T_0)$ for several samples $\mathbf r^N_0$, temperatures $T$, and sources $\epsilon$. The colorbar encodes the relative distance of the couple $(T,~\epsilon)$ used for the simulation from $\epsilon^*(T; \mathbf r^N_0,T_0)$ in the $\epsilon$ direction. All data collapse on a master curve which is well fitted by Eq.~(\ref{eqn:tau_chi}) (dashed line), with $c={0.015(2)}$, $\tau_0={88(13)}$ and $z={1.15(10)}$. The inset shows a tentative power-law fit which obviously fails at high values of the susceptibility. (b) Disorder-averaged overlap autocorrelation function along the Widom line for  several temperatures. The full lines represent a fit to the empirical form presented in the main text, with $C_0\approx 0.54$ and $\phi\approx 8.2$.}
\label{fig:dynamics}
\end{figure}

\section{Critical dynamics}
We now turn to the study of the dynamics in the vicinity of the critical point, a study which has never been attempted before. Relaxation on approaching a critical point is characterized by a slowing down and a divergence of the relaxation time exactly at criticality. In the case of the RFIM, the slowing down is anomalous and described by an activated dynamic scaling according to which it is not the relaxation time $\tau(T)$ that grows as a power law of the correlation length $\xi(T)$, as usual, but its logarithm. In a renormalization-group framework this reflects the property that criticality is controlled by a zero-temperature fixed point~\cite{villain1985equilibrium,fisher1986scaling}. From the correlation function of the overlap in Eq.~(\ref{eqn:CQQ}) we define a relaxation time $\tau(\epsilon,T; \mathbf r_0^N,T_0)$ as the time at which $C(t|\epsilon,T;\mathbf r^N_0,T_0) = 0.2$. We approach the critical point from above and consider points ($T$, $\epsilon$) at or close to the Widom line. Instead of the correlation length $\xi(T)$ to which we do not have direct access we use the connected susceptibility $\chi_\mathrm{con}$ which scales as $\xi^{2-\eta}$. For the $3d$ RFIM, $2-\eta\approx \theta$ and it has further been shown that $\psi=\theta$~\cite{balog2015activated} (so that $\xi^\psi\sim \chi_\mathrm{con}$). We therefore consider the following form~\cite{parmar1994dynamic},
\begin{equation}
\tau(\epsilon,T;\mathbf r_0^N,T_0)=\tau_0 [\chi_T(\epsilon,T;\mathbf r_0^N,T_0)]^{z/\theta}e^{c\,\chi_T(\epsilon,T;\mathbf r_0^N, T_0)},
\label{eqn:tau_chi}
\end{equation}
with $\tau_0$ and $c$ some constants and $z$ a dynamical exponent describing some subdominant behavior. Whereas the dominant activated scaling behavior is independent of the dynamics (the overlap is in any case a nonconserved order parameter), the subdominant behavior and prefactors can be somehow modified by choosing an appropriate algorithm. Here, we consider the swap algorithm that is expected to speed up any pre-asymptotic dynamics. (We find that the ordinary Monte Carlo dynamics is much too slow near the critical point.) Fig.~\ref{fig:dynamics}~(a) shows that the data agree well with the prediction in Eq.~(\ref{eqn:tau_chi}). The increase in relaxation time is limited to a little more than two orders of magnitude but it is sufficient to distinguish between activated scaling (main panel) and conventional power-law scaling (inset). 

Another prediction of the activated dynamic scaling in the RFIM is that the correlation function should be very stretched, on a logarithmic scale, with $C(t;T)=\widetilde{C}(\ln t/\ln \tau(T))$~\cite{fisher1986scaling} and $\widetilde{C}(x)$ a scaling function for which no theoretical prediction is available. We find that along the Widom line, we can fit our autocorrelation data $C^*(t|T;T_0)=\overline{C(t|\epsilon,T;\mathbf r^N_0,T_0)}$ with an empirical form previously used in RFIM-like systems~\cite{ogielski1986critical,dierker1987random,valiullin2006exploration}, $\widetilde C(x)=C_0 \exp(-x^\phi)$, with $C_0$ and $\phi$ two $T$-independent adjustable parameters. As seen in Fig.~\ref{fig:dynamics}~(b), data at large times for all temperatures agree well with this prediction. A rescaling using the variable $t/\tau$ is instead inconsistent with the data. We stress that the activated critical slowing down that we analyze here in the vicinity of the critical point at ($T_c$, $\epsilon_c$) is unrelated to the glassy slowdown of the bulk glass-former, but requires the existence of a critical point in the RFIM universality class. 

\begin{figure}
\includegraphics[width=4.27cm]{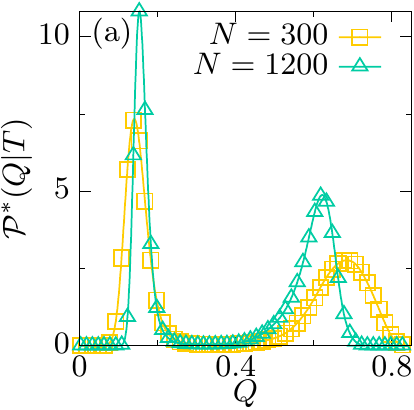}
\includegraphics[width=4.27cm]{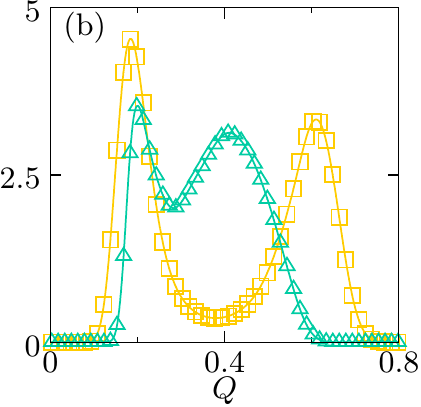}
\caption{Disorder-averaged overlap probability distribution below [$T=0.085$ in (a)] and above [$T=0.100$ in (b)] the critical point for the case $T_0=T$.}
\label{fig:proba_T}
\end{figure}

\section{Influence of the temperature of the reference configuration}
We finally come back to the issue of the persistence of a critical point when the temperature of the reference configuration is $T_0=T$. This situation then probes typical states of the landscape and can be more directly related to the physics of a glass-forming liquid with no applied source. As already stressed, such study is computationally more demanding. We have therefore limited ourselves to checking the existence of a transition, without studying its nature in detail nor investigating the critical dynamics. The results are illustrated in Fig.~\ref{fig:proba_T} where we show the disorder-averaged overlap probability distribution $\mathcal{P}^*(Q\vert T)$. For $T=0.085$ [in (a)], it becomes increasingly bimodal with $N$, the reduced half-width of the low overlap peak shrinking with $N$ consistently with the existence of a first-order transition. By contrast, for $T=0.100$ [in (b)], the probability function is bimodal at small $N$ but the peaks rapidly approach each other as $N$ increases, indicating that $\mathcal{P}^*(Q\vert T)$ should become single-peaked in the thermodynamic limit. Overall, our results suggest that the critical point also exists when $T_0=T$, with $0.085\leq T_c < 0.100$, close to or below the mode-coupling crossover, as eluded by past studies~\cite{cammarota2010phase,berthier2015evidence,berthier2017configurational}.

\section{Conclusions}
To sum up, we have performed an extensive finite-size scaling analysis of a critical point proposed to characterize three-dimensional glass-formers, relying on the massive speedup afforded by the swap Monte Carlo algorithm combined with umbrella sampling techniques. \rev{Our results demonstrate for the first time the existence in the thermodynamic limit} of a critical point, with a first-order transition line at lower temperatures, and our finite-size scaling analysis is consistent with the RFIM universality class in three dimensions. The critical point studied here is unique, since it represents, to date, the only piece of the mean-field/RFOT theoretical construction to survive other than as a crossover the introduction of finite-dimensional fluctuations. \rev{This closes, for three-dimensional liquids, a 25-year-old quest since its initial analysis in a fully mean-field context and more recent field-theoretical predictions,} and gives us hope that a fundamental understanding of glass formation can be further developed in finite dimensions. 

\begin{acknowledgments}
We thank G. Biroli, C. Cammarota, D. Coslovich, M. Ediger, and R. Jack for fruitful discussions. Some simulations were performed at MESO@LR-Platform at the University of Montpellier. B. Guiselin acknowledges support by Capital Fund Management - Fondation pour la Recherche. This work was supported by a grant from the Simons Foundation (Grant No. 454933, L.B.).
\end{acknowledgments}

\bibliographystyle{apsrev4-1}
\bibliography{./biblio.bib}

\end{document}


\title{Supplemental Material: Random-field Ising model criticality in a glass-forming liquid}

\author{Benjamin Guiselin}%

\affiliation{Laboratoire Charles Coulomb (L2C), Universit\'e de Montpellier, CNRS, 34095 Montpellier, France}

\author{Ludovic Berthier}%

\affiliation{Laboratoire Charles Coulomb (L2C), Universit\'e de Montpellier, CNRS, 34095 Montpellier, France}

\affiliation{Department of Chemistry, University of Cambridge, Lensfield Road, Cambridge CB2 1EW, United Kingdom}

\author{Gilles Tarjus}%
\affiliation{LPTMC, CNRS-UMR 7600, Sorbonne Universit\'e, 4 Pl. Jussieu, F-75005 Paris, France}

\date{\today}

\maketitle

\setcounter{equation}{0}
\setcounter{figure}{0}
\setcounter{table}{0}
\setcounter{page}{1}
\renewcommand{\theequation}{S\arabic{equation}}
\renewcommand{\thefigure}{S\arabic{figure}}
\renewcommand{\bibnumfmt}[1]{[S#1]}
\renewcommand{\citenumfont}[1]{S#1}
\renewcommand\thesubsection{\arabic{subsection}}
\renewcommand\thesubsubsection{\arabic{subsection}.\arabic{subsubsection}  }


\section{Methods}

\label{sec:methods}

The system under study is a three-dimensional continuously polydisperse mixture of $N$ spherical particles of equal mass $m$ in a cubic box with periodic boundary conditions. Diameters $\sigma$ are drawn from the distribution $p(\sigma)\propto\sigma^{-3}$ for $\sigma \in [\sigma_\mathrm{min};\ \sigma_\mathrm{max}]$. The potential energy contribution to the hamiltonian reads $\mathcal{H}[\mathbf r^N]=~\sum_{1\leq i<j\leq N}v(|r^{(i)}- r^{(j)}|/\sigma_{ij})$ with the interparticle potential $v(x)=v_0(1/x^{12}+c_0+c_2 x^2+c_4 x^4)$ for $x<x_\mathrm{cut}=1.25$ and 0 otherwise~\cite{gutierrez2015static}. The constants $c_0$, $c_2$ and $c_4$ are chosen so that the potential and its two first derivatives are continuous at $x_\mathrm{cut}$. Cross-diameters are nonadditive to avoid crystallization and demixing~\cite{ninarello2017models}: $\sigma_{ij}=0.5\left(\sigma_i+\sigma_j\right)\left(1-0.2\left\vert\sigma_i-\sigma_j\right\vert\right)$. Energies are expressed in units of $v_0$, lengthscales in units of the average diameter $\mu=\int_{\sigma_\mathrm{min}}^{\sigma_\mathrm{max}}\mathrm{d}\sigma p(\sigma)\sigma$ and timescales in units of $\sqrt{m\mu^2/v_0}$. With this convention, $\sigma_\mathrm{min} = 0.726$ and $\sigma_\mathrm{max}=1.6095$. The number  density $\rho=N/L^3$ equals $1$. To define the overlap, we choose $w(x)=e^{-\ln(2)x^4}$ and $a=0.22$ but results are qualitatively insensitive to these choices. 

Simulations are performed using a hybrid scheme~\cite{berthier2019efficient} combining molecular dynamics (MD) in the canonical ensemble with a Nos\'e-Hoover thermostat~\cite{hoover1985canonical} and swap Monte Carlo moves to faster the relaxation~\cite{ninarello2017models} and thus the sampling of configuration space. The hybrid scheme relies on the repetition of the following procedure~\cite{berthier2019efficient}. First, MD is run during $n_{MD}=25$ timesteps. We use a Liouville-based reversible integrator~\cite{martyna1996explicit,frenkel2001understanding} with a timestep $dt=0.01$ and a damping time of the thermostat $\tau_\mathrm{th}=0.5$~\cite{martyna1992nose}. Then, MD is stopped, positions and velocities of particles are frozen and $N$ elementary swap moves are intented. In an elementary swap move, the diameters of two particles chosen randomly are exchanged~\cite{ninarello2017models} and this move is accepted following Metropolis criterion~\cite{allen2017computer,frenkel2001understanding}. Swap moves fulfil detailed balance, ensuring a complete sampling in the $NVT$-ensemble.

First, equilibrium configurations at $T_0=0.06\lesssim T_g$ are obtained by equilibrating the system. After equilibration, the swap $\alpha$-relaxation time $\tau_{\alpha,{\rm swap}}$ is measured, defined when the self-intermediate scattering function equals $e^{-1}$. Eventually, a long run is performed and independent reference configurations are stored every $2\tau_{\alpha,{\rm swap}}$. 

For the thermodynamic study, a harmonic bias is added to the hamiltonian of the liquid in order to force the system to visit atypical configurations, 
\begin{equation}
\mathcal{H}_{Q_0}[\mathbf r^N; \mathbf r_0^N]=\mathcal{H}[\mathbf r^N]+\frac{Nk}{2}(\hat{Q}[\mathbf r^N;\mathbf r_0^N]-Q_0)^2,
\label{eqn:hamiltonian_US}
\end{equation}
a technique known as umbrella sampling~\cite{frenkel2001understanding,torrie1977monte} \rev{(US)} or Gaussian ensemble sampling~\cite{challa1988gaussian,costeniuc2005generalized}. The spring constant $k$ controls the fluctuations of the overlap around $Q_0$ and we fix $k=20$ to make them narrow and nearly Gaussian. For each temperature $T$ and quenched reference configuration $\mathbf r_0^N$, we run around 30 simulations with different values of $Q_0$ to cover the entire range of overlap values. For a given value of $Q_0$, equilibration is checked by running two simulations, one initiated from the reference configuration and one from an equilibrium configuration at temperature $T$, and by checking that both time series of $\hat{Q}$ converge to the same average value and that the mean squared displacement exceeds $1.5$.

\rev{The measure for each $Q_0$ of the most probable value $\tilde{Q}(Q_0\vert T; \mathbf r_0^N, T_0)$ of the overlap enables us to compute the derivative with respect to $Q$ of the Franz-Parisi potential at temperature $T$ and for a given reference configuration $\mathbf r_0^N$
\begin{equation}
V'(Q\vert T; \mathbf r_0^N, T_0) = k(Q_0- Q) \hspace*{0.3cm}\text{ for }\hspace*{0.3cm} Q=\tilde{Q}(Q_0\vert T; \mathbf r_0^N, T_0).
\label{eqn:derivative_FP}
\end{equation} 
This equality can be derived as follows. First, we use the relation between the Franz-Parisi potential and the probability distribution of the overlap in the bulk liquid 
\begin{equation}
\mathcal{P}(Q\vert T; \mathbf r_0^N, T_0)\propto e^{-N\beta V(Q\vert T; \mathbf r_0^N, T_0)},
\label{eqn:def_FP}
\end{equation}
with a proportionality constant which depends on $T$, $T_0$ and $\mathbf r_0^N$ [see Eq.~(1) of the main text]. Then, under the sampling with hamiltonian~(\ref{eqn:hamiltonian_US}), the probability distribution of the overlap yields 
\begin{equation}
\mathcal{P}_{(k,Q_0)}(Q\vert T; \mathbf r_0^N, T_0)\propto\mathcal{P}(Q\vert T; \mathbf r_0^N, T_0)e^{-N\beta k(Q-Q_0)^2/2}\propto e^{-N\beta [V(Q\vert T; \mathbf r_0^N, T_0)+k(Q-Q_0)^2/2]},
\label{eqn:proba_US}
\end{equation}
the last equality resulting from Eq.~(\ref{eqn:def_FP}). In Eq.~(\ref{eqn:proba_US}), the proportionality constant now depends on $Q_0$, $k$, $T$, $T_0$ and $\mathbf r_0^N$. In order to get rid of this normalisation constant, we differentiate the last equation with respect to $Q$ and we evaluate it at the most probable value of the overlap $Q=\tilde{Q}(Q_0\vert T; \mathbf r_0^N, T_0)$ [at the maximum of $\mathcal{P}_{(k,Q_0)}(Q\vert T; \mathbf r_0^N, T_0)$]. Thus, the left-hand side vanishes and we are left with Eq.~(\ref{eqn:derivative_FP}).}

\rev{Thanks to the swap algorithm, thermalisation is achieved for each US simulation even without parallel tempering~\cite{hukushima1996exchange}. As a result, our method does not require overlap between probability distributions for adjacent values of $Q_0$. In the simulations, we have thus chosen a large value of $k$ in order for the $\tilde{Q}(Q_0\vert T; \mathbf r_0^N, T_0)$'s to be well-defined with small uncertainties. Besides, the method can be implemented for arbitrary-large systems.}

The derivative of $V(Q\vert T; \mathbf r_0^N, T_0)$ evaluated at all \rev{$Q=\tilde{Q}(Q_0\vert T; \mathbf r_0^N, T_0)$}'s is then interpolated using a cubic spline interpolation and integrated  to obtain the Franz-Parisi potential~\cite{fernandez2009tethered,press2007numerical}. Finally, thermodynamic properties for any field $\epsilon$ are deduced from the Boltzmann distribution of hamiltonian $\mathcal{H}_\epsilon$ (see main text). \rev{This amounts in computing the first cumulants of the probability distribution of the overlap at temperature $T$ and coupling field $\epsilon$ for a given reference configuration $\mathbf r_0^N$, the latter can be written as
\begin{equation}
\mathcal{P}(Q\vert \epsilon,T;\mathbf r_0^N,T_0)\propto e^{-N\beta[V(Q\vert T; \mathbf r_0^N, T_0)-\epsilon Q]},
\end{equation}
up to a normalisation constant that can be computed directly.} The entire procedure is eventually repeated for different $T$ and $\mathbf r_0^N$.

To study the dynamics, we run simulations using swap with the hamiltonian $\mathcal{H}_\epsilon$ for 25 different samples $\mathbf r_0^N$. We approach the critical point from above, using several couples $(T,~\epsilon)$ close to or at the Widom line determined from the thermodynamic study. The liquid is first equilibrated and one equilibrium configuration $\mathbf r^N (\epsilon,T;\mathbf r_0^N,T_0)$ is stored. Equilibration is checked similarly as in the static study. Then, a long run is performed, starting from configuration $\mathbf r^N (\epsilon,T;\mathbf r_0^N,T_0)$, to compute $C(t|\epsilon,T;\mathbf r_0^N,T_0)$ and deduce the autocorrelation time $\tau(\epsilon,T;\mathbf r_0^N,T_0)$ when the autocorrelation equals 0.2. We self-consistently check afterward that the simulation lasted at least 10 times the autocorrelation time. In a second time, we also look at the average over disorder of the autocorrelation function along the Widom line, ${C}^*(t|T;T_0)=\overline{C(t|\epsilon,T;\mathbf r_0^N,T_0)}$. As this quantity requires a good sampling of the correlation function for each reference configuration $\mathbf r_0^N$ separately, we run 60 simulations in the iso-configurational ensemble starting from the very same configuration $\mathbf r^N (\epsilon,T;\mathbf r_0^N,T_0)$ with initial velocities drawn from the Boltzmann distribution at temperature $T$~\cite{widmer2004reproducible,wansleben1991monte,pearson1985dynamic}. 

In addition, to compute the thermal susceptibility $\chi_T(\epsilon,T;\mathbf r_0^N,T_0)$ , we have taken advantage of the thermodynamic calculations at temperature $T=0.22$ and $T=0.25$ and then extrapolated to temperatures $T=0.23$ and $T=0.26$ using a reweighting in temperature. In short, the Franz-Parisi potential $V(Q\vert T_\mathrm{e}; \mathbf r_0^N, T_0)$ at a target temperature $T_\mathrm{e}$ close to $T$ is derived from the Franz-Parisi potential at temperature $T$ by using the following formula:
\begin{equation}
V(Q\vert T_\mathrm{e}; \mathbf r_0^N, T_0) = \frac{T_\mathrm{e}}{T}V(Q\vert T; \mathbf r_0^N, T_0) +\frac{1-T_\mathrm{e}/T}{N}E_1(Q\vert T; \mathbf r_0^N, T_0)
-\frac{(1-T_\mathrm{e}/T)^2}{2NT_\mathrm{e}} E_2(Q\vert T; \mathbf r_0^N, T_0)-\frac{T_\mathrm{e}}{2N}\ln E_2(Q\vert T; \mathbf r_0^N, T_0),
\label{eqn:FP_extrapolT}
\end{equation}
where $E_n(Q\vert T; \mathbf r_0^N, T_0)$ stands for the $n$th cumulant of the sum of the kinetic and the potential energies at fixed value of $Q$. In the course of umbrella simulations, these quantities evaluated at the most probable value of the overlap $\hat{Q}(Q_0\vert T; \mathbf r_0^N, T_0)$ are directly measured, and then interpolated to any overlap value using a cubic spline interpolation.

\rev{To derive Eq.~(\ref{eqn:FP_extrapolT}), we introduce $\mathcal{P}(E,Q\vert T;\mathbf r_0^N, T_0)$ the joint probability of total energy $E$ and overlap $Q$ with the reference configuration $\mathbf r_0^N$ in the bulk liquid at temperature $T$. We then assume that the conditional probability of $E$ given an overlap value $Q$ can reasonably be represented by a gaussian with mean $E_1(Q\vert T; \mathbf r_0^N, T_0)$ and variance $E_2(Q\vert T; \mathbf r_0^N, T_0)$ so that the joint probability yields
\begin{equation}
\mathcal{P}(E,Q\vert T; \mathbf r_0^N, T_0)=\mathcal{P}(Q\vert T; \mathbf r_0^N, T_0)\times[2\pi E_2(Q\vert T; \mathbf r_0^N, T_0)]^{-1/2} e^{-[E-E_1(Q\vert T; \mathbf r_0^N, T_0)]^2/[2E_2(Q\vert T; \mathbf r_0^N, T_0)]}.
\end{equation}
As we kept track of the total energy, we can thus reweight the joint probability at any target temperature $T_\mathrm{e}$ [$\beta_\mathrm{e}=1/(k_B T_\mathrm{e})$]: $\mathcal{P}(E,Q\vert T_\mathrm{e}; \mathbf r_0^N, T_0)\propto \mathcal{P}(E,Q\vert T; \mathbf r_0^N, T_0) e^{-(\beta_\mathrm{e}-\beta)E}$. Consequently, by integrating over energies, we can extrapolate the probability distribution of the overlap in the bulk liquid at temperature $T_\mathrm{e}$. Eq.~(\ref{eqn:FP_extrapolT}) then follows from the definition of the Franz-Parisi potential [see Eq.~(\ref{eqn:def_FP})].}

\section{FSS analysis at the first-order transition}

\begin{figure}[!h]
\includegraphics[width=4.27cm]{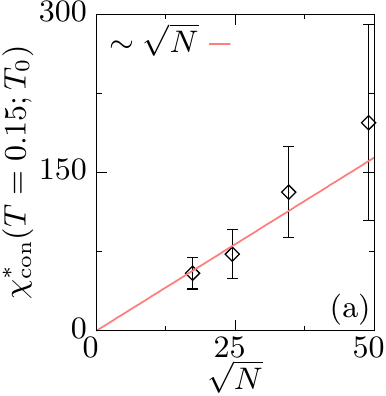}
\includegraphics[width=4.27cm]{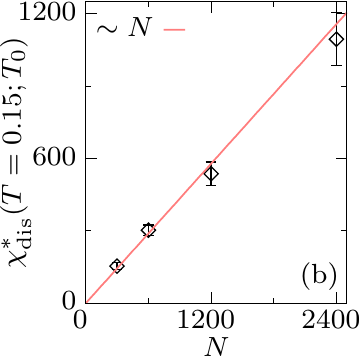}
\includegraphics[width=4.27cm]{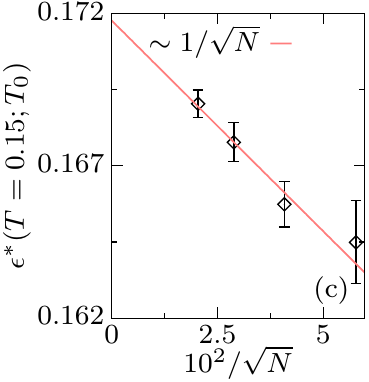}
\includegraphics[width=4.27cm]{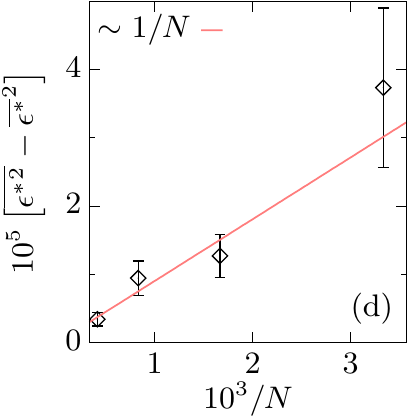}
\caption{Finite-size scaling analysis in the first-order transition region ($T=0.15$) for the maximum connected susceptibility (a), the maximum disconnected susceptibility (b), the average over disorder of the value of $\epsilon$ at coexistence (c) and its variance (d). Uncertainties are computed with the jacknife method~\cite{newman1999monte}.}
\label{fig:FSS_1st_order}
\end{figure}

In Fig.~\ref{fig:FSS_1st_order}, we present further evidence of the existence of a first-order transition at low enough temperature in the thermodynamic limit. We show that as far as susceptibilities are concerned, they display the expected scaling $\chi_\mathrm{con}^*(T=0.15;T_0)\sim \sqrt{N}$ and $\chi_\mathrm{dis}^*(T=0.15;T_0)\sim N$ for RFIM-like systems. In addition, $\epsilon^*(T=0.15;T_0)$ converges to its thermodynamic limit value as $1/\sqrt{N}$ while the variance of $\epsilon^*(T=0.15;\mathbf r_0^N,T_0)$ vanishes as $1/N$~\cite{vink2008colloid,vink2010finite}.

\section{FSS analysis close to the critical point}

\begin{figure}[!h]
\includegraphics[width=4.27cm]{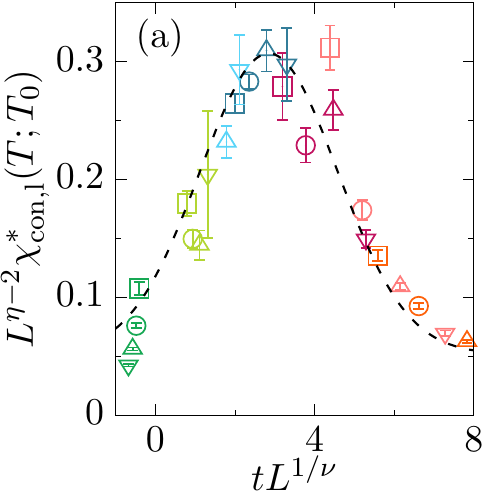}
\includegraphics[width=4.27cm]{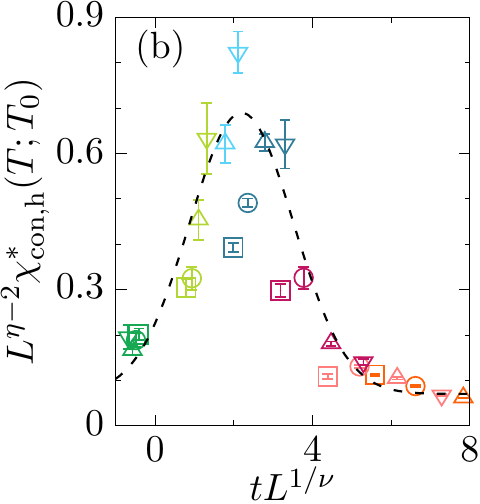}
\caption{Connected susceptibility in the low-overlap phase (a) and in the high-overlap phase (b) versus reduced temperature, following the FSS ansatz proposed in the main text. The dashed line is a guide for the eye. Uncertainties are computed with the jacknife method.}
\label{fig:FSS_SI}
\end{figure}

In Fig.~\ref{fig:FSS_SI}, we show further evidence for the existence of a RFIM critical point at $T_c\approx 0.167$. Following the FSS ansatz presented in the main text, we display the rescaled connected susceptibility in the low-overlap and high-overlap phases. Data from all system sizes collapse reasonably well. Data for the connected susceptibility in the high-overlap phase are more scattered, though. From sample to sample, the position and the width of the high-overlap peak of the probability distribution both fluctuate much more strongly than the low-overlap peak's counterparts. Thus, it is likely that a much larger number of independent reference configurations would be required to obtain a better convergence of $\chi_\mathrm{con,h}$.

\bibliography{./biblio.bib}